\documentclass[showpacs,twocolumn,prl]{revtex4}
\usepackage{epsf}
\usepackage{graphicx}
\date{\today}
\title{Results and discussion}

\newcommand{\br}{{\mathbf r}}

\newcommand{\bx}{{\mathbf x}}

\begin{document}
\title{Universal Scaling of the Conductivity at the Superfluid-Insulator
Phase Transition}
\author{Jurij \v{S}makov and Erik S\o rensen}
\affiliation{Department of Physics and Astronomy, McMaster University, 
Hamilton, Ontario L8S 4M1, Canada}
\begin{abstract}
The scaling of the conductivity at the superfluid-insulator quantum phase
transition in two dimensions is studied by numerical simulations of
the Bose-Hubbard model. In contrast to previous studies, we focus on
properties of this model in the experimentally relevant thermodynamic
limit at finite temperature $T$. We find clear evidence for {\it
deviations} from $\omega_k$-scaling of the conductivity towards
$\omega_k/T$-scaling at low Matsubara frequencies $\omega_k$. By
careful analytic continuation using Pad\'e approximants we show that
this behavior carries over to the real frequency axis where the
conductivity scales with $\omega/T$ at small frequencies and
low temperatures. We estimate the universal dc conductivity to be
$\sigma^\star=0.45(5)Q^2/h$, distinct from previous estimates
in the $T=0$, $\omega/T\gg 1$ limit.
\end{abstract}

 \maketitle The non-trivial
properties of materials in the vicinity of quantum phase
transitions~\cite{QP} (QPTs) are an object of intense
theoretical~\cite{QP,sondhi,DamleSachdev} and experimental
studies. The effect of quantum fluctuations driving the QPTs is
especially pronounced in low-dimensional systems, such as
high-temperature superconductors and two-dimensional (2D) electron
gases, exhibiting the quantum Hall effect. Particularly valuable are
theoretical predictions of the behavior of the dynamical response
functions, such as the optical conductivity and the dynamic structure
factor, since they allow for direct comparison of the theoretical
results with experimental data. It was pointed out by Damle and
Sachdev~\cite{DamleSachdev}, that at the quantum-critical coupling the
scaled dynamic conductivity $T^{(2-d)/z}\sigma(\omega,T)$ at low
frequencies and temperatures is a function of the single variable
$\hbar\omega/k_BT$:
\begin{equation}
\label{sachdev_scaling}
\sigma(\omega/T,T\to0)=\left({k_BT}/{\hbar c}\right)^{(d-2)/z}\sigma_{Q}\,\Sigma(\hbar\omega/k_BT).
\end{equation}
Here $\sigma_Q=Q^2/h$ is the conductivity ``quantum'' ($Q=2e$ for the
models we consider), $\Sigma(x\equiv\hbar\omega/k_BT)$ is a universal
dimensionless scaling function, $c$ a non-universal constant, and $z$
the dynamical critical exponent. For $d=2$ the exponent
vanishes, leading to a purely universal conductivity~\cite{fisher90a},
depending only on frequency $\omega$, measured against a
characteristic time $\hbar\beta$ set by finite temperature $T$
as $\hbar\omega/k_BT$. Once $\hbar\omega/k_BT\gg 1$, for {\it fixed}
$T$, the system no longer ``feels'' the effect of finite temperature
and it is natural to expect that at such high $\omega$ a crossover to
a temperature-independent regime will take place~\cite{sondhi}, so
that $\sigma(\omega,T)\sim\sigma(\omega)$ with $\sigma(\omega)$
decaying at high frequencies as $1/\omega^2$ \cite{DamleSachdev}.
Deviations from scaling of $\sigma$ with $\omega$ therefore signal
that temperature effects have become important.  Note that the
predicted universal behavior occurs for {\it fixed} $\omega/T$ as
$T\to 0$.
The physical mechanisms of transport are predicted~\cite{DamleSachdev}
to be quite distinct in the different regimes determined by the value
of the scaling variable $x$: hydrodynamic, collision-dominated for
$x\ll 1$, and collisionless, phase-coherent for
$x\gg1$ with $\Sigma=\Sigma(\infty)$ largely independent of $x$ in $d=2$ 
and $\sigma$ independent of $T$~\cite{DamleSachdev,sachdev}.
\begin{figure}[t]
\includegraphics*[width=0.9\hsize,scale=1.0]{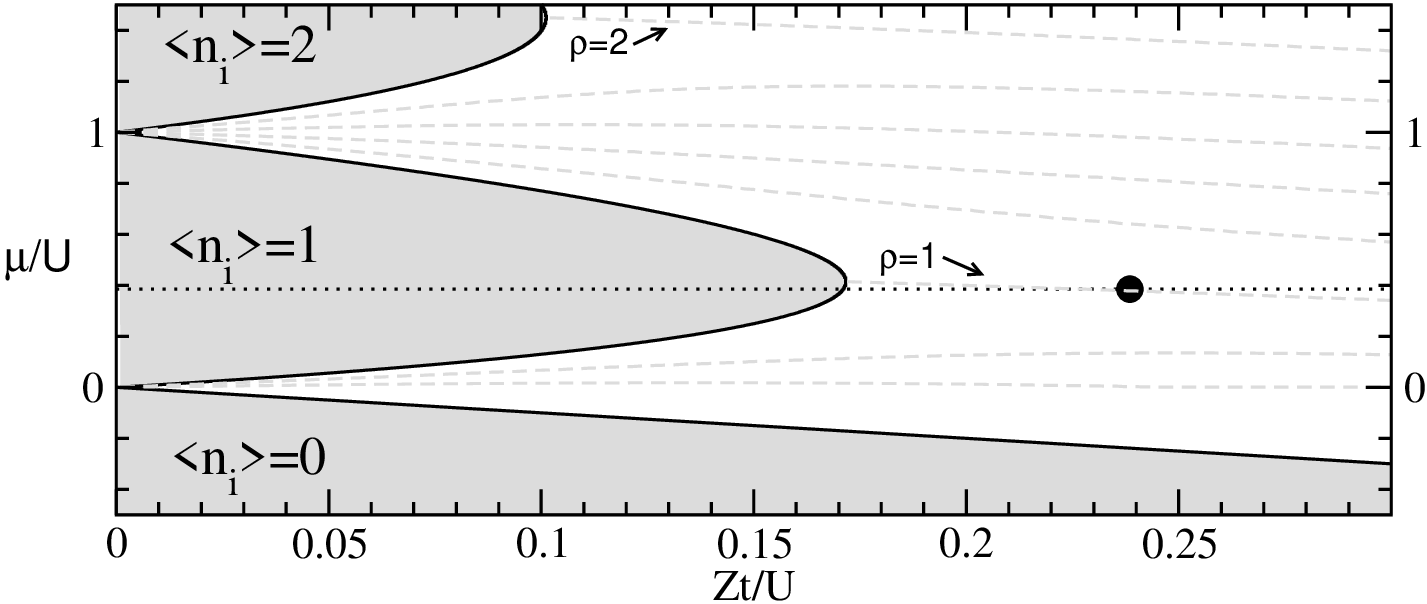}
\caption{Mean-field ground state phase diagram of the 2D Bose-Hubbard
model. Shaded areas are the Mott insulating phases.
Dashed lines are constant density profiles in steps of $0.2$. $\bullet$
indicates the location of QCP at the tip of Mott lobe as determined by
SSE simulations along the dotted line.
\label{fig:pdiag}
}
\end{figure}

Intriguingly, early numerical studies~\cite{cha, sorensen, blstw, sth}
of QPTs in model systems have failed to observe scaling with
$\hbar\omega/k_BT$. The results of the experiments seeking to verify
the scaling hypothesis are ambiguous as well. Some of them, performed
at the 2D quantum Hall transitions~\cite{Engel} and 3D metal-insulator
transitions~\cite{Lee}, appear to support it. Others either note the
absence of scaling~\cite{Balaban} or suggest a different scaling
form~\cite{yslee}. While the discrepancy between theory and experiment
may be attributed to the unsuitable choice of the measurement
regime~\cite{DamleSachdev}, typically leading to $\hbar\omega/k_BT \gg
1$, there is no good reason why the predicted scaling would not be
observable in numerical simulations if careful extrapolations first to
$L\to\infty$ and then $T\to 0$  for fixed $\omega/T$ are performed.

Our primary goal is to resolve this controversy by performing precise
numerical simulations of the frequency-dependent conductivity at
finite temperatures in the vicinity of the 2D QPT, exploiting recent
algorithmic advances to access larger system sizes and a wider
temperature range. After the extrapolation of the results to the
thermodynamic and $T=0$ limits and careful analytic continuation, we
are able to demonstrate how the predicted universal behavior of the
conductivity may indeed be revealed.

We consider the 2D Bose-Hubbard model with the Hamiltonian ${\mathcal
H}_{\rm BH}={\mathcal H}_0+{\mathcal H}_1$, where the first term
describes the non-interacting softcore bosons hopping via the
nearest-neighbor links of a 2D square lattice, and the second one
includes the Hubbard-like on-site interactions:
\begin{eqnarray}
\label{kin_ham}
{\mathcal H}_0&=&-t\sum_{\br,\delta} (b^{\dagger}_{\br}b_{\br+\delta}+
b^{\dagger}_{\br+\delta}b_{\br})-\mu\sum_{\br}n_{\br},\\
{\mathcal H}_1&=&\frac{U}{2}\sum_{\br}n_{\br}(n_{\br}-1).
\end{eqnarray}
Here $\delta={\bf x,y}$, $n_{\br}=b^{\dagger}_{\br}b_{\br}$ is the
particle number operator on site $\br$, and $b^{\dagger}_{\br}, b_{\bf
r}$ are the boson creation and annihilation operators at site $\br$.
Model parameters include the hopping constant $t$, Hubbard repulsion
$U$, and chemical potential $\mu$. The mean-field ground state phase
diagram of this model (Fig. \ref{fig:pdiag}) displays a number of
Mott-insulating lobes with fixed integer boson density at low $Zt/U$
($Z=4$ is the lattice coordination number). As the hopping $t$ is
increased or $\mu$ is varied, a QPT to a superfluid (SF) phase takes
place. We
concentrate on the QPT occuring at
the tip of the Mott lobe along the path of constant $\mu$,
distinct from the generic transition occurring elsewhere along the
phase boundary~\cite{fisher}.

The numerical simulations of ${\mathcal H}_{\rm BH}$ were performed
using the stochastic series expansion (SSE) technique with directed
loop updates~\cite{old_sandvik,dirloop}, which is known to be very
efficient for the simulations of boson models. Furthermore,
as described below, it allows us to directly evaluate the relevant
correlation functions without discretization or numerical
integration over the imaginary time.
We have also employed an alternative $(2+1)$-dimensional classical
representation~\cite{fisherlee,sorensen} of ${\mathcal H}_{\rm BH}$ in
terms of link-current variables describing the total 
bosonic current ${\bf J}=(J^x,J^y,J^\tau)$ defined on a discrete
$L\times L\times L_\tau$ space-time lattice
($L_\tau\Delta_\tau=\hbar\beta$):
\begin{equation}
{\mathcal H}_V=\frac{1}{K} {\sum_{({\bf r},\tau)}}
\left[\frac{1}{2} {\bf J}_{({\bf r},\tau)}^{2}- \mu J_{({\bf r},\tau)}^\tau\right].
\label{eq:hV}
\end{equation}
${\mathbf J}$ has to be conserved and is therefore divergence-free,
${\bf \nabla \cdot J} = 0$.  The link-current variables take on
integer values $J^{x,y,\tau}=0,\pm 1, \pm 2, \ldots$ and denote the
deviation of the particle number from its mean, so the transition
corresponds to $\mu=0$. 
$K$ is the effective temperature, varying like $t/U$ in
${\mathcal H}_{\rm BH}$. The model defined by ${\mathcal H}_V$ has
been studied in the past using a very efficient directed geometrical
worm algorithm~\cite{GWorms}, and its critical point at $\mu=0$ has
been determined~\cite{GWorms} to be $K_c=0.33305(5)$. A drawback of
this representation is that the time direction is discrete, imposing
an ultra-violet frequency cut-off of data at $\omega_c=1/\Delta\tau$,
in contrast to SSE, where there is no such problem. The two numerical
approaches are therefore largely complementary. These advanced
techniques allowed us to simulate lattices of linear sizes up to
$L=30$ and inverse temperatures up to $\beta=10$ using SSE for
${\mathcal H}_{\rm BH}$, and $L=256$, $L_{\tau}=64$ using the directed
geometrical worm algorithm for ${\mathcal H}_V$.

Performing SSE simulations of ${\mathcal H}_{\rm BH}$ we first
precisely locate the quantum critical point
(QCP) for fixed $\mu/U$ at the tip of the first Mott lobe. This
transition belongs to the (2+1)D $XY$ universality class with a
dynamical critical exponent $z=1$~\cite{fisher}.  In the vicinity of
the QCP the SF density $\rho_s$ and compressibility $\kappa$ are
expected to obey the scaling relations
\begin{eqnarray}
\label{sfscaling}
\rho_s&=&L^{2-d-z}Y_1(\delta L^{1/\nu}, \beta L^{-z}),\\
\label{kappascaling}
\kappa&=&L^{z-d}Y_2(\delta L^{1/\nu}, \beta L^{-z}).
\end{eqnarray}
Here $\nu$ is the correlation length critical exponent,
$\beta=1/k_BT$, $\delta=|t-t_c|$, and $Y_{1,2}(x,y)$ are the
two-variable scaling functions.
\begin{figure}[t]
\includegraphics*[angle=0,width=\hsize,scale=1.0]{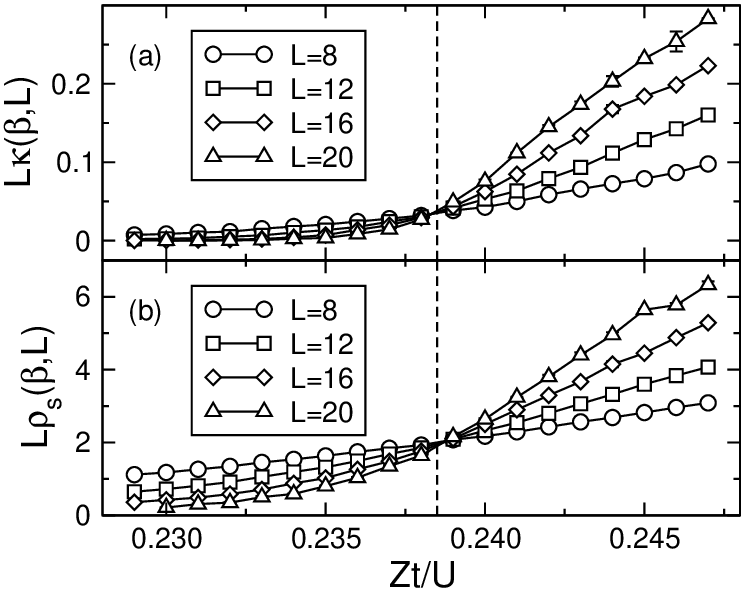}
\caption{The compressibility (a) and SF density (b) versus $Zt/U$
at fixed $\mu/U=0.375$ and aspect ratio $\beta L^{-z}=0.5$. 
The dashed vertical line is drawn at $Zt_c/U=0.2385$.
Error bars are displayed only if larger than the
symbol size.
\label{fig:crit_cross}
}
\end{figure} 
For a fixed aspect ratio $\beta L^{-z}$ plots of the $L\rho_s$ and
$L\kappa$ should then intersect at the critical point
$\delta=0$. Results of such a calculation at constant $\mu_c/U=0.375$
are presented in Fig. \ref{fig:crit_cross}, from which we determine
$Zt_c/U=0.2385(5)$, $\nu=0.66(5)$.
The position of the QCP and the shape of the phase boundary in its
vicinity is consistent with previous simulations \cite{krauth} and
strong-coupling perturbation theory \cite{monien}. 

To analyze the behavior of the zero-momentum conductivity in the
vicinity of the QCP we employ the relation between the dynamic
conductivity $\sigma(\omega)$ and the Fourier transform
$\Lambda_{xx}(\omega)$ of the time-dependent current-current
correlation function (CCCF) established by the Kubo formula
\cite{mahan,swz}. In SSE the real-time CCCF required to determine
$\Lambda_{xx}(\omega)$ is not directly accessible. Instead, the
standard approach is to measure the CCCF $\Lambda_{xx}(\tau)=\langle
j_x(\tau)j_x(0)\rangle$ on the imaginary time axis, calculate its
Fourier transform $\Lambda_{xx}(i\omega_k)$ as a function of the
Matsubara frequencies $\omega_k\equiv 2\pi k/\beta$, and analytically
continue the result to real frequencies~\cite{swz, blstw, sth}. Here
and below we adopt a unit system in which both $Q$ and $\hbar$ are
unity, and $j_x(\tau)$ is the Heisenberg representation of the current
operator
$j_x=it[b^{\dagger}_{\br+\bx}b_{\br}-b^{\dagger}_{\br}b_{\br+\bx}]$.
We have:
\begin{equation}
\label{kubo}
\sigma(i\omega_k)=2\pi \sigma_Q\frac{\langle -k_x\rangle-\Lambda_{xx}(i\omega_k)}{\omega_k}
\equiv 2\pi\sigma_Q\frac{\rho(i\omega_k)}{\omega_k}.
\label{eq:cond}
\end{equation}
Here $\langle -k_x\rangle$ is the kinetic energy per link and
$\rho(i\omega_k)$ is the frequency-dependent stiffness. To measure
$\Lambda_{xx}(i\omega_k)$ we note that $\Lambda_{xx}(\tau)$
may be expressed in terms of the correlation functions
$\Lambda^{\gamma\nu}_{xx}(\br,\tau)=\langle
K^{\gamma}_x(\br,\tau) K^{\nu}_x({\mathbf 0}, 0)\rangle$ of operators
$K_{x}^{+}(\br,\tau)=t\,b^{\dagger}_{\br+\bx}(\tau)b_{\br}(\tau)$
and $K_{x}^{-}(\br,\tau)=t\,b^{\dagger}_{\br}(\tau)b_{\br+\bx}(\tau)$,
which may be estimated efficiently in SSE~\cite{old_sandvik}. Remarkably, 
it is possible to analytically perform the Fourier transform with respect to
$\tau$ yielding
\begin{equation}
\label{lambda_hyper}
\Lambda^{\gamma\nu}_{xx}(\br,\omega_k)=
\left\langle\frac{1}{\beta}\sum_{m=0}^{n-2}\, \bar{a}_{mn}(\omega_k)
N(\nu,\gamma; m)\right\rangle,
\end{equation}
where $N(\nu, \gamma; m)$ is the number of times the operators
$K^{\gamma}(\br)$ and $K^{\nu}({\bf 0})$ appear in the SSE operator
sequence separated by $m$ operator positions, and $n$ is the
expansion order. The coefficients $\bar{a}_{mn}(\omega_k)$ are given
by the degenerate hypergeometric (Kummer) function:
$
\bar{a}_{mn}(\omega_k)={}_1F_1(m+1, n;-i\beta\omega_k).
$
This expression and (\ref{lambda_hyper}) allow us to
directly evaluate $\Lambda_{xx}(\br,\omega_k)$ as a function of
Matsubara frequencies,
eliminating any errors associated with the
discretization of the imaginary time interval.  Analogously, in the
link-current representation $\rho(i\omega_k)$ can be
calculated~\cite{sorensen}, and the conductivity can be obtained from
Eq.~(\ref{eq:cond}).
\begin{figure}[b] 
\includegraphics*[angle=0,width=8.6cm]{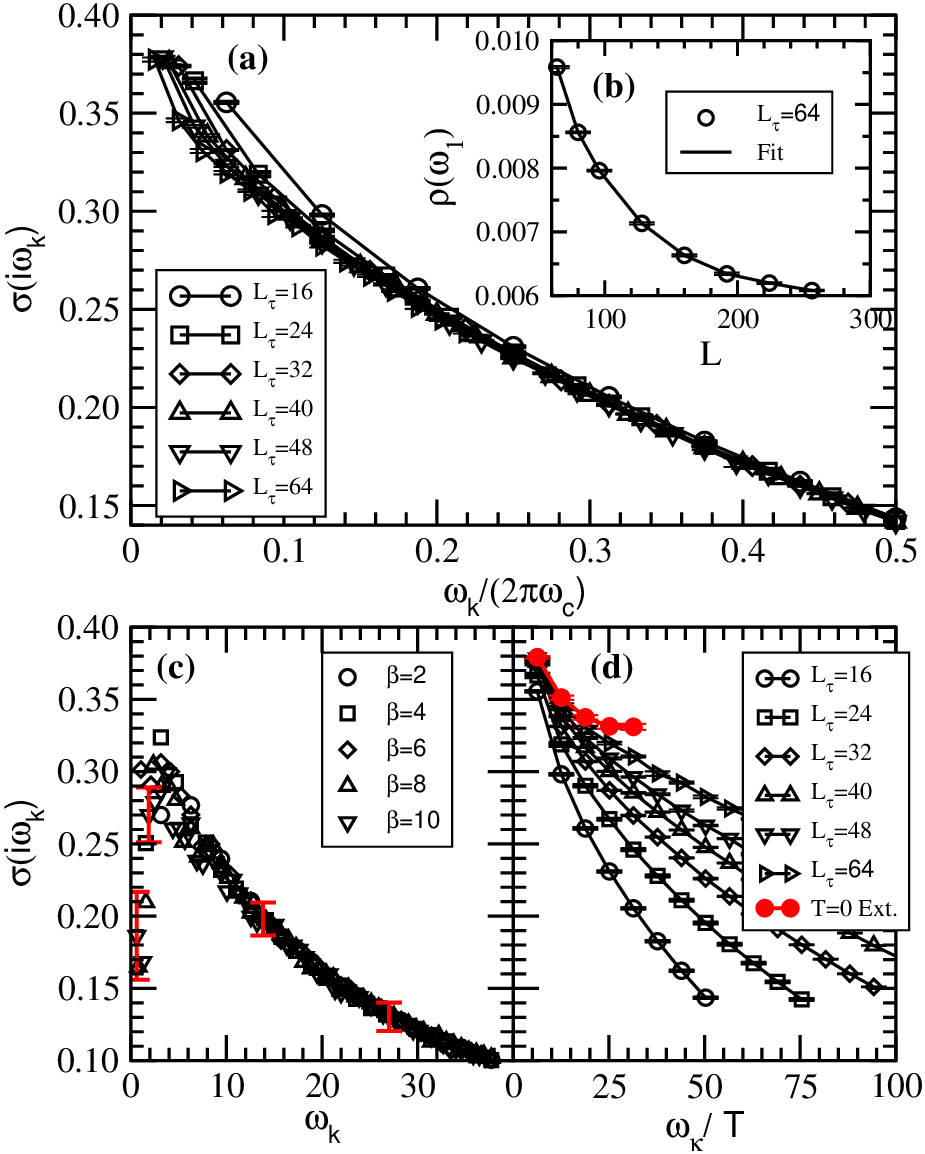}
\caption{The conductivity $\sigma(\omega_k)$ in units of $\sigma_Q$
versus Matsubara frequency $\omega_k/\omega_c$ as obtained from
${\mathcal H}_{\rm V}$ (a). All results have been extrapolated to the
thermodynamic limit $L\to\infty$ using the scaling form 
$f(L)=a+b\exp(-L/\xi)/\sqrt{L}$~\cite{privman}
by calculating $\rho(\omega_k)$ at
fixed $L_\tau$ using 9 lattice sizes from $L=L_\tau\ldots 4L_\tau$ as
shown in (b). $\sigma(\omega_k)$ in units of $\sigma_Q$ versus
Matsubara frequency $\omega_k$ as obtained from SSE calculations of
${\mathcal H}_{\rm BH}$, with some typical error bars shown. All
results have been extrapolated to the thermodynamic limit by
calculating $\Lambda_{xx}(\omega_k)$ for fixed $\beta$ using 5 lattice
sizes $L=12\ldots 30$ (c). Scaling plot of the conductivity data from (a)
versus $\omega_k/T$. $\bullet$ denotes extrapolations to $T\to0$
($L_\tau\to\infty$)
at fixed $\omega_k/T$
using: $f(L_\tau)=c+d\exp(-L_\tau/\xi_\tau)/\sqrt{L_\tau}$~\cite{privman} (d).
\label{fig:blambda}
}
\end{figure}

\begin{figure*}[t] 
\includegraphics*[angle=0,width=0.85\textwidth,scale=0.9]{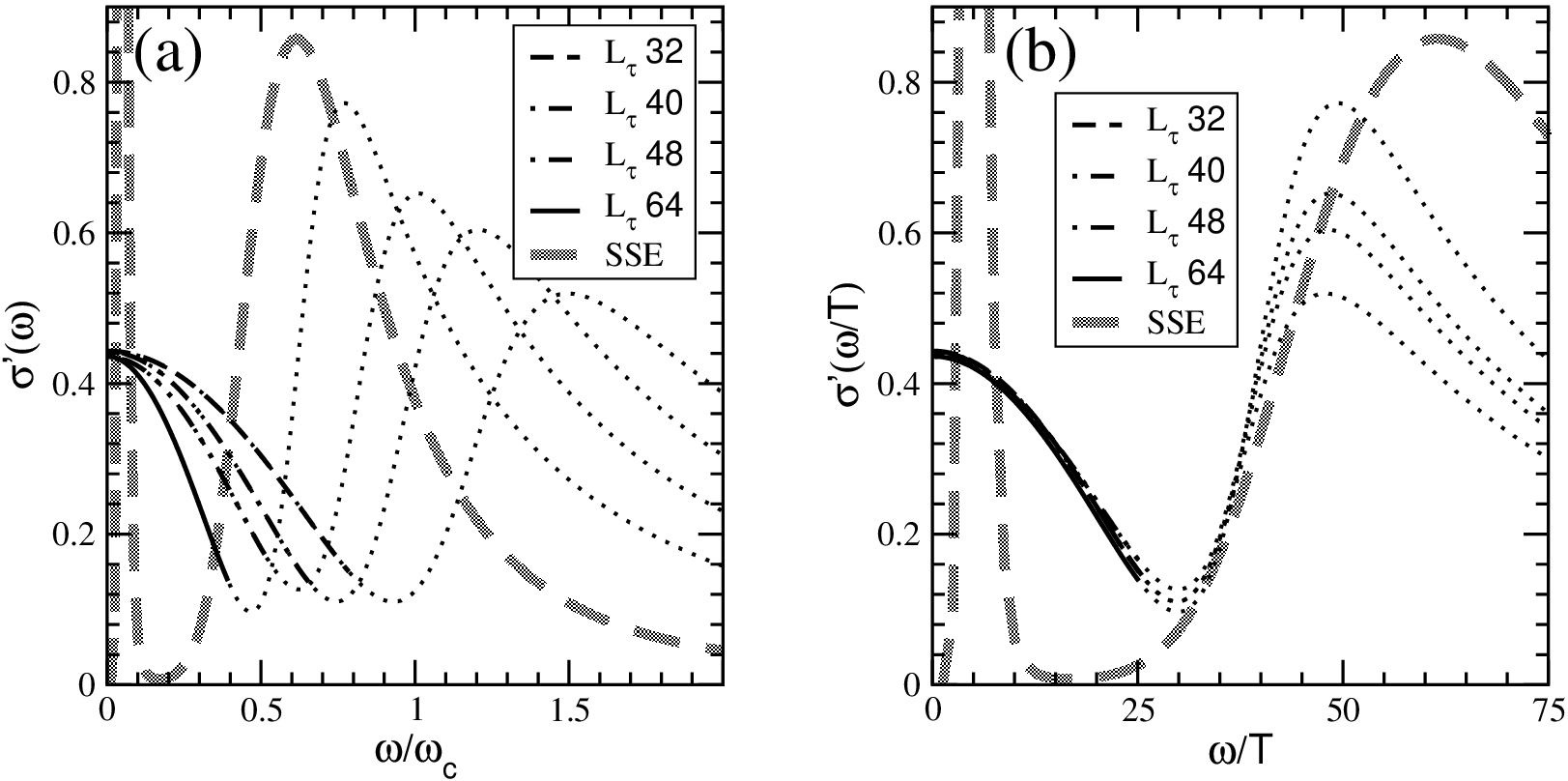}
\caption{The real part of the conductivity $\sigma'$ at the
critical coupling in units of $\sigma_Q$. The data marked $L_\tau$, plotted versus $\omega/\omega_c$,
were obtained using ${\mathcal H}_{V}$, combined with the analytic
continuation of $\rho(\omega/\omega_c)$ as explained in the
text. Results for $\omega/\omega_c\gtrsim 1/2$ are denoted by dotted lines.
The data marked SSE, plotted versus $\omega/10$, were obtained by direct SSE simulations of
${\mathcal H}_{\rm BH}$ with $L=20$, $\beta=10$ and subsequent maximum
entropy analysis (a).  Results as a function of $\omega/T$ (b).
\label{fig:cont}
}
\end{figure*} 

In Fig.~\ref{fig:blambda} we show results for $\sigma(i\omega_k)$
versus $\omega_k$ obtained using the geometrical worm algorithm on
${\mathcal H}_V$ at $K_c$ (Fig.~\ref{fig:blambda}a) and by SSE
simulations at $t_c,\mu_c$ of ${\mathcal H}_{\rm BH}$
(Fig.~\ref{fig:blambda}c). In both cases the results have been
extrapolated to the thermodynamic limit $L\to\infty$ at fixed
$\beta$. As evident from Fig.~\ref{fig:blambda}a, the results {\it
deviate} from scaling with $\omega_k$ at small $\omega_k$ and more
significantly so at higher temperatures (small $L_\tau$).  These
deviations are also visible in the continuous time SSE data in
Fig.~\ref{fig:blambda}c, demonstrating that they cannot be attributed
to time discretization errors. Similar deviations have been noted
previously~\cite{cha,sorensen} but were {\it not} analyzed at fixed
$\beta$. Since the deviations persist in the $L\to\infty$ limit at
fixed $\beta$, they may only be interpreted as {\it finite T}
effects. Expecting a crossover to $\omega_k/T$ scaling at small
$\omega_k$, we plot our results versus $\omega_k/T$ in
Fig.~\ref{fig:blambda}d.  For $L_\tau\ge 32$, $\sigma(\omega_1/T)$ is
already {\it independent} of $T$ ($L_\tau$). In fact, as shown in
Fig.~\ref{fig:blambda}d, for $\omega_{1\ldots 5}$,
$\sigma(\omega_k/T,T)$ can unambigously be extrapolated to a {\it
finite} $\sigma(\omega_k/T,T\to 0)\sim\Sigma(x)$ limit.  This fact is
a clear indication that $\omega_k/T$-scaling indeed occurs as $T\to
0$.  Tentatively, for increasing $\omega_k/T$,
$\sigma(\omega_k/T,T\to0)$ appears to reach a constant value of
roughly $0.33\sigma_Q\sim\Sigma(\infty)$ in excellent agreement with
theoretical estimates~\cite{DamleSachdev,Fazio}.  We note that
deviations from $\omega_k$-scaling appear to be largely {\it absent} in
simulations of ${\mathcal H}_{\rm BH}$ with
disorder~\cite{sorensen,blstw}. However, at this QCP the dynamical critical
exponent is different ($z=2$). As is evident from the size of the
error bars in Fig.~\ref{fig:blambda}, simulations of ${\mathcal H}_V$
are much more efficient than the SSE simulations directly on
${\mathcal H}_{\rm BH}$.  In the following analytic continuation we
therefore use the SSE data mostly as a consistency check.

Our results on the imaginary frequency axis are limited by the lowest
Matsubara frequency, $\omega_1=2\pi k_BT/\hbar$. However, the
information about the behavior of $\sigma'(\omega)\equiv
\textrm{Re}\,\sigma(\omega)$ at low $\omega$ is embedded in values of
the CCCF at all Matsubara frequencies, allowing us to determine it. In
order to study the $\omega/T$-scaling predicted for the hydrodynamic
collision dominated regime~\cite{DamleSachdev} $\hbar\omega/k_BT\ll 1$,
we have attempted analytic continuations of $\rho(i\omega_k)$ to
obtain $\sigma'(\omega)$ at real frequencies. SSE results for
${\mathcal H}_{\rm BH}$ were analytically continued using the Bryan
maximum entropy (ME) method~\cite{jarrell} with flat initial
image. For the results obtained for the link-current model ${\mathcal
H}_V$ we use a method that should be most sensitive to low frequencies
$\omega/\omega_c<1$ or $\Sigma(x\ll 1)$. We fit the extrapolated low frequency part (first
$10$-$15$ Matsubara frequencies) of $\rho(i\omega_k)$ to a 6th-order
polynomial. The resulting 6 coefficients are then used to obtain a
$(3,3)$ Pad\'e approximant using standard techniques~\cite{numrec}.
This approximant is then used for the analytic continuation of $\rho$ by
$i\omega_k\to \omega+i\delta$.  Resulting real frequency
conductivities $\sigma'(\omega)$ are displayed in Fig. \ref{fig:cont}a
versus $\omega/\omega_c$. The typical SSE data are plotted versus
$\omega/10$ and are only shown for $L=20$, $\beta=10$). 

The results for ${\mathcal H}_V$ show a broadened peak as $\omega\to
0$, due to inelastic scattering, followed by a second peak nicely
consistent in height and width with the SSE data. The SSE data also
displays a high narrow peak at very low frequencies, whose position
and shape are unstable with respect to the choice of the initial image
and MaxEnt parameters. This is clearly an artifact of the method,
however its presence is indicative of the tendency to accumulate the
weight at very low frequencies, in qualitative agreement with
${\mathcal H}_V$ result. The subsequent fall-off in the conductivity
at high frequencies is physically consistent, but its functional form
depends on the Pad\'e approximant used.  {}For $\omega/\omega_c\gtrsim
1/2$, we expect the analytic continuation of the data for ${\mathcal
H}_V$ to become sensitive to the order of the approximant used and
we therefore indicate the results in this regime by dotted lines only.
We note that results at {\it all} temperatures yield the same dc
conductivity $\sigma^\star=0.45(5)\sigma_Q$, theoretically
predicted~\cite{fisher90a} to be universal.  Due to the very different
scaling procedure this result differs from previous numerical result
$\sigma^\star=0.285(20)\sigma_Q$ on the same model~\cite{cha} in the
$T=0$ limit. It also differs significantly from a theoretical
estimate~\cite{DamleSachdev}, $\sigma^\star=1.037\sigma_Q$, valid to
leading order in $\epsilon=3-d$. 
Remarkably, our result for the dc conductivity is very close to the
one obtained in Ref.~\onlinecite{blstw} for the phase transition in the
disordered Bose-Hubbard model.
Experimental results indicate a value close to unity~\cite{goldman},
however it was previously observed~\cite{sorensen} that long-range
Coulomb interactions, impossible to include in the present study, tend
to increase $\sigma$ considerably.
The same data are shown versus
$\omega/T$ in Fig.~\ref{fig:cont}b. Notably, when using this
parametrization $\omega_c$ {\it cancels out} and all our data follow
the {\it same} functional form. The scaling with $\omega/T$ at low
frequencies is now immediately apparent, with a surprisingly wide low
$\omega/T$  peak.  The width of this peak is consistent with the data
in Fig.~\ref{fig:blambda}d. Furthermore, on the same $\omega/T$ scale the
continuous time SSE data for ${\mathcal H}_{BH}$ and the results for
${\mathcal H}_V$ qualitatively agree. 

In summary, we have demonstrated that by doing a very careful data analysis it
is possible to observe the theoretically predicted universal
$\omega/T$-scaling at the 2D superfluid-insulator transition. We have
also estimated the universal dc conductivity at this transition and
found that it differs significantly from existing numerical and
theoretical estimates.

We thank S.~Sachdev, S.~Girvin and A.~P.~Young for valuable comments
and critical remarks. Financial support from SHARCNET, NSERC and CFI
is gratefully acknowledged. All calculations were done at the SHARCNET
facility at McMaster University.

\end{document}